\documentclass[aps,pre,onecolumn,11pt,notitlepage,nofootinbib]{revtex4-1}
 \pdfoutput=1
\usepackage{setspace}
\usepackage{multirow}
\usepackage[dvipsnames]{xcolor}
\usepackage{epsfig,graphics,amssymb,amsmath,amsthm,color,setspace}
\usepackage{graphicx,txfonts}
\usepackage{booktabs}
\usepackage[sfdefault=cmbr]{isomath}

\newtheorem*{theorem}{Theorem 1}

\begin{document}

\title{{There's more than one way to cancel a regularized Stokeslet}}
\author{William H. Mitchell and Dona Pantova}
\affiliation{Macalester College, 1600 Grand Avenue, St. Paul, MN, 55105, USA}
\date{\today}

\begin{abstract}
The Green's functions of Stokes flow are widely used in the analysis and simulation of microscale fluid flows. We adapt a procedure from H.A. Lorentz for the method of images in Stokes flow to the regularized setting. Our solutions differ from those previously reported, a surprising result given the uniqueness theory for elliptic partial differential equations. The discrepancy originates in the fact that the two versions are exact solutions of inhomogeneous Stokes systems with slightly different forcing on the right-hand sides.  We compare the fluid flows produced by the two methods and conclude that the Lorentz versions may be advantageous in some settings. 
\end{abstract}

\maketitle

\section{Introduction}
\begin{figure}[h!]
\begin{center}
\includegraphics[width=\linewidth]{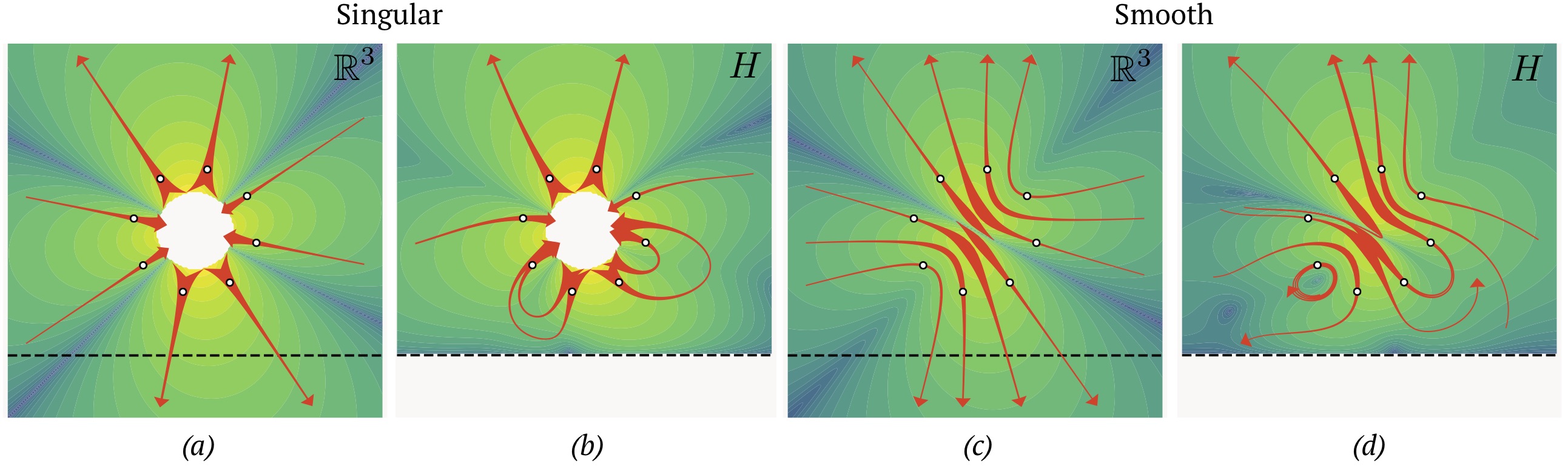}
\end{center}
\caption{Four different stresslet flow fields. The dashed line indicates the $xy$-plane; trajectories can pass through it when the flow domain has no boundary ($ac$) but not when the flow domain is bounded by a no-slip wall at $\{z=0\}$ ($bd$). In the leftmost two panels ($ab$), the velocity becomes unbounded at the source point in the center of the frame and these flows are accordingly called \emph{singular}. In contrast, the \emph{regularized} flows in the rightmost two panels ($cd$) are smooth everywhere. In all cases, arrow thickness indicates fluid velocity and the trajectories are integrated forwards and backwards from the eight seed points indicated by dots. The contour fields in the background also indicate velocity magnitude, with color plotted on a log scale. These are two-dimensional slices of three-dimensional flows, so a trajectory can approach the wall as $t\to-\infty$ as on the right side of ($d$) without contradicting incompressibility.  
The \emph{Lorentz reflection theorem} gives a procedure for producing a flow on the half-space $H$ from a flow on $\mathbb{R}^3$, e.g. for producing the flow fields $(b)$ and $(d)$ from $(a)$ and $(c)$. The object of this paper is to describe the wall-bounded regularized flows obtained through Lorentz' construction, of which the stresslet $(d)$ is one example.  The formulas we obtain are not equivalent to those given by Ainley and Cortez, a surprising result which we discuss in Sec. \ref{sec:comp}.  
}
\label{fig:stresslets}
\end{figure}

The linearity of the Stokes equations of fluid flow permits the use of fundamental solutions, also known as Green's functions. Typically these are defined as solutions of an inhomogeneous Stokes system of differential equations with singular forcing functions, leading to infinite velocities at the source points (see Fig. \ref{fig:stresslets}ac). For some applications this is an undesirable property numerically.  This issue can be addressed through the use of regularized Stokeslets; see \cite{Fauci2005} and references therein.  A regularized fundamental solution remains finite even at the source point and is defined as the solution of a Stokes system with smooth forcing in either the force balance or the continuity equation (see Fig. \ref{fig:stresslets}bd). For settings involving an infinite plane wall, the method of images is an attractive technique because it avoids discretizing the planar boundary. One set of images for the regularized Green's functions was given previously using an ad hoc method \cite{adebc08,cortez2015general}; in this paper we obtain a different set of images by generalizing a procedure described by H.A. Lorentz more than a century ago.  This procedure gives velocity and pressure fields that, like the Cortez systems, exactly cancel the free-space velocity on a planar boundary; however, the two versions are not identical. After discussing Lorentz's procedure and applying it in the case of regularized flow, we compare our systems to those previously known.  We find that the Lorentz versions may be preferable in some contexts, but the choice of the correct regularization parameter for a given problem is more important than the choice between the Cortez systems and the Lorentz versions presented here. 

\begin{table}
\begin{tabular}{ccccc}
Decay rate &\,\quad\,& Blob $\phi(r)$ &\qquad& Companion Blob $\phi^d(r)$\\
&&&&\\
Algebraic && $\displaystyle\frac{15\delta^4}{8\pi(r^2+\delta^2)^{7/2}}$ && $\displaystyle  \frac{3\delta^2}{4\pi(r^2+\delta^2)^{5/2}}$\\
&&&&\\
Exponential && $\displaystyle \frac{5\delta^2+5\delta r - r^2}{64\pi\delta^5}\exp(-r/\delta) $ && $\displaystyle \frac{r+\delta}{32\pi\delta^4}\exp(-r/\delta) $\\
\end{tabular}
\caption{Four blob functions, reproduced from Table 1 of \cite{cortez2015general} with a correction: the denominator in the first exponentially decaying blob has $\delta^5$ instead of $\delta^4$. The argument $r$ is in turn a scalar field on $\mathbb{R}^3$, \emph{e.g.} $r = r(\bm x) = |\bm x-\bm y|$ where $\bm y$ is the \emph{source point} and $\bm x$ is the \emph{observation point.}  If $\psi$ is any of the four blob functions printed here, then $\psi(|\bm x-\bm y|)$ has unit mass on $\mathbb{R}^3$ and the \emph{regularization parameter} $\delta$ controls how concentrated this mass is around $\bm y$.}
\label{tbl:Blobs}
\end{table}

\begin{figure}
\begin{center}
\includegraphics[width=0.5\linewidth]{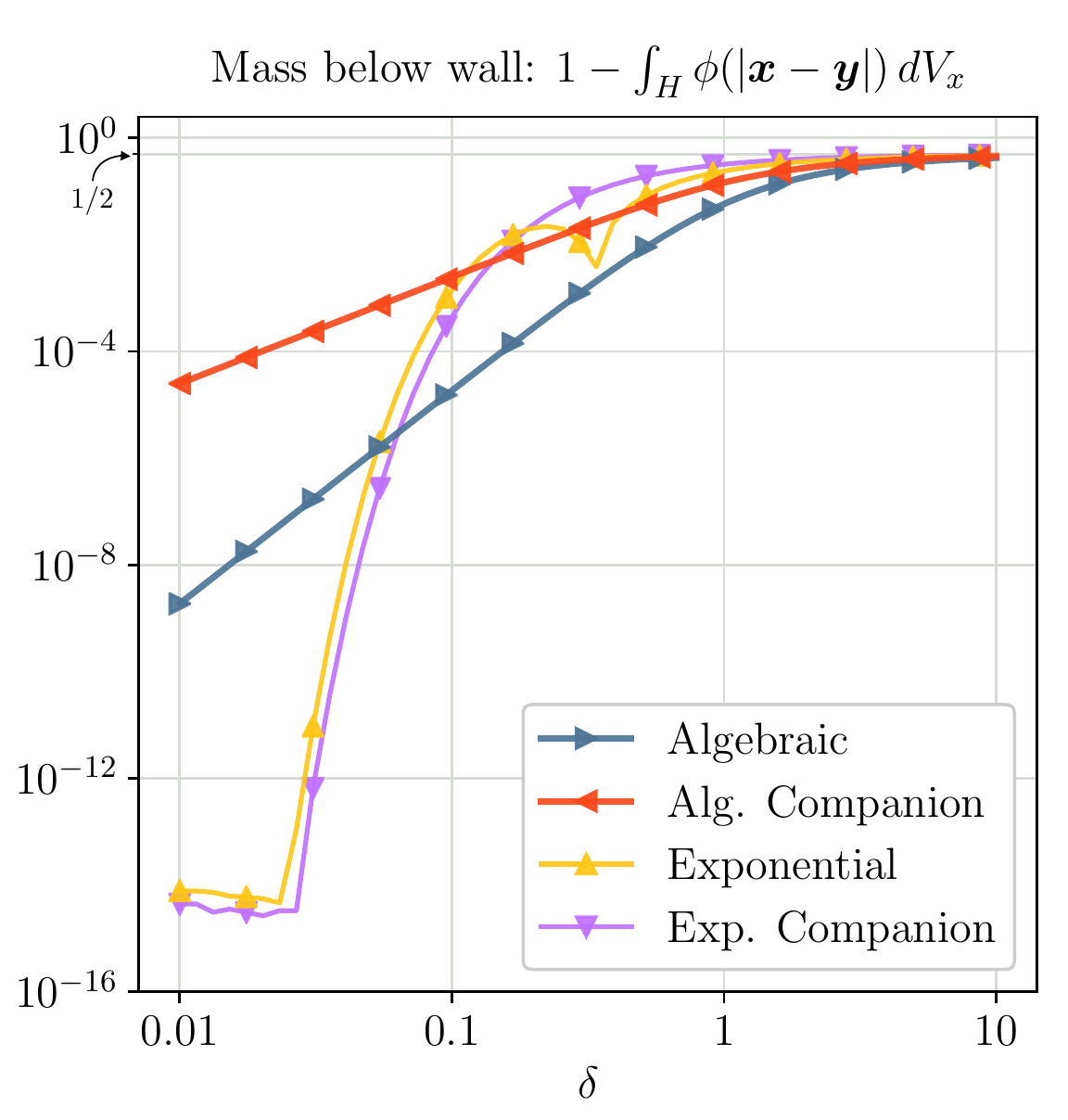}
\end{center}
\caption{The amount of mass lying below the plane $\{x_3=0\}$, for several choices of regularized delta functions centered at $\bm y=(0,0,1)$.  
When $\delta \ll 1$, the mass is concentrated around $\bm y$ and so the amount falling below the wall is small, especially for the exponentially decaying blobs.  When $\delta$ becomes larger, the mass below the wall increases towards $\frac12$. Within both the exponential and algebraic blob families, we see slower decay of the companion blobs relative to the standard ones. In section \ref{sec:comp} we argue that this slower decay is a reason to avoid the companion blobs entirely, as in Lorentz's version of the method of images. } 
\end{figure}
\section{The Lorentz Reflection Theorem}
More than a century ago, H.A. Lorentz gave a procedure for finding an image flow $(\bm u^*,p^*)$ for a given free-space Stokes flow $(\bm u,p)$ such that $(\bm u^*,p^*)$ is also a Stokes flow and $\bm u+\bm u^*$ vanishes on the wall \cite{lorentz1896,kuiken1996ha}. The theorem has been used several times to obtain the image systems for singular Green's functions; to our knowledge it has not previously been applied for the regularized case. In this section we give details and examples for both settings.  
\subsection{The LRT for singular flows}
The procedure is as follows.\footnote{Descriptions of Lorentz's reflection procedure have appeared in several places \cite{lorentz1896,kk91,kuiken1996ha}. Our presentation is based on the discussion in Kim and Karilla but we write $\bm v$ and $\bm u^*$ instead of defining the hat and star operators.} We start with a velocity $\bm u$ and pressure $p$ satisfying $\partial_i u_i = 0$ and $-\partial_i p + \partial_m\partial_m u_i = 0$ on $\mathbb{R}^3 \setminus\{\bm y\}$, where $y_3>0$. Let $\beta_{ij} = \delta_{ij}-2\delta_{3i}\delta_{3j}$ denote reflection through the wall and define 
\begin{align}
\label{eq:pro1}
 v_i&= -\beta_{ij}u_j - 2x_3 \partial_i u_3 + x_3^2 \partial_m\partial_m u_i \\
\label{eq:pro2} q&=p + 2x_3 \partial_3p - 4\partial_3 u_3\\
\label{eq:pro3} u^*_i &= \beta_{ij}v_j(\bm\beta\cdot\bm x) \\
 p^*&= q(\bm\beta\cdot\bm x). \label{eq:pro4}
 \end{align}
The corrected velocity and pressure are $\bm U = \bm u+\bm u^*$ and $P = p + p^*$. The corrected velocity vanishes on the wall at $\{x_3=0\}$. Moreover, the pair $(\bm U, P)$ solves the same PDE as $(\bm u,p)$ above the wall, that is, $\bm U$ $\partial_iU_i = 0$ and $-\partial_i p + \partial_m\partial_m u_i = 0$ hold at each point above the wall other than $\bm y$, and $\bm U$ has the same singular behavior as $\bm u$ at $\bm y$ since $\bm u^*$ is smooth above the wall. 
\subsubsection{Example: the image system for the singular point source}
If fluid is injected into an three-dimensional domain with no boundaries at a point $\bm y$, then the flow at $\bm x$ is given by 
\[\bm u_i(\bm x) = \frac{x_i-y_i}{4\pi|\bm x-\bm y|^3} = \frac{X_i}{4\pi R^3}\]
where $\bm X = \bm x-\bm y$ and $R = |\bm X|$. The accompanying pressure field is constant. In this nondimensional formulation, one unit of volume is created at $\bm y$ during one unit of time.  
If $\bm y_3>0$, we can ask how this $\bm u$ needs to be modified to accommodate a no-slip boundary at $\{x_3=0\}$. In contrast to the Laplace setting, it is not enough to place another point source at the reflection point $\beta_{ij}y_j = (y_1,y_2,-y_3)$; this cancels the normal but not the tangential component of velocity on the wall.  Instead, we follow Lorentz' procedure \eqref{eq:pro1}-\eqref{eq:pro4}, using the Einstein summation convention to streamline the calculus operations.  In the first stage we obtain the pair $(\bm v,q)$ given by 
\begin{align}
v_i &= \frac{-\beta_{ij}X_j- 2 \delta_{3i}x_3}{4\pi R^3}+ \frac{3 x_3 X_i X_3}{2\pi R^5}\\
q &= \frac{-1}{\pi R^3}+ \frac{3X_3^2}{\pi R^5} 
\end{align}
Next we obtain $p^*$ from $q$ by substituting $\bm\beta\cdot\bm x$ for $\bm x$, or equivalently substituting $-x_3$ for $x_3$.  Defining $\hat{X_i} = \beta_{ij}x_j-y_i$ and $\hat R = \left|\hat{\bm X}\right|$, we have 
\begin{equation}
p^* = \frac{-1}{\pi \hat R^3}+ \frac{3\hat X_3^2}{\pi \hat R^5}.
\end{equation}
Similarly, we follow \eqref{eq:pro4} to obtain $\bm u^*$ from $\bm v$ by replacing $x_3$ with $-x_3$ and then reflecting:
\begin{equation}
u_i^* = \beta_{ij}\left(\frac{-\beta_{jk}\hat X_k+ 2 \delta_{3j}x_3}{4\pi \hat R^3}- \frac{3 x_3 \hat X_j \hat X_3}{2\pi \hat R^5}\right) 
=\frac{-\hat X_i- 2 \delta_{3i}x_3}{4\pi \hat R^3}- \frac{3 x_3  \beta_{ij}\hat X_j \hat X_3}{2\pi \hat R^5}
\end{equation}
The result can be slightly condensed by noting that $\hat X_i + 2\delta_{3i}x_3 = X_i$. The result is 
\begin{align}
U_i &= \frac{X_i}{4\pi R^3} -\frac{X_i}{4\pi \hat R^3}- \frac{3 x_3  \beta_{ij}\hat X_j \hat X_3}{2\pi \hat R^5}\\
P &= 0 - \frac{1}{\pi \hat R^3}+ \frac{3\hat X_3^2}{\pi \hat R^5}
\end{align}
The corrected solution $(\bm U,P)$ satisfies $\bm U|_{x_3=0}=\bm 0$.  Moreover, at all $\bm x$ above the wall and distinct from $\bm y$ we have $\partial_iU_i=0$ and $-\partial_i P + \partial_m\partial_m U_i = 0$. That is, the forcing in the Stokes equations has not changed during the passage from $(\bm u,p)$ to $(\bm U, P)$. The situation will be different for the regularized analogue, and we consider this situation next.  

\subsection{The LRT for regularized flows}
When one or both of the continuity and force balance equations have nonzero forcing over an open region instead of just at a point, the Lorentz procedure still produces an image system $(\bm u^*,p^*)$ with the property that $\bm U=\bm u+\bm u^*$ vanishes on the wall.  However, the image system is a Stokes flow with nonzero forcing and so the PDE satisfied by the corrected flow has to be modified. 
We now present a theorem identifying the PDE which is solved by the flow produced by Lorentz' procedure when applied in the regularized setting.  
\begin{theorem}
Suppose that $\phi$ is a smooth scalar field and $\bm \psi$ is a smooth vector field on $\mathbb{R}^3$ and that these fields are the right-hand sides of a Stokes system:
\begin{align}
\partial_i u_i(\bm x) &= \phi(\bm x)\\
-\partial_i p(\bm x) + \partial_m\partial_m u_i(\bm x) &= \psi_i(\bm x).
\end{align}
Let $\bm u^*$ and $p^*$ be defined by \eqref{eq:pro1}-\eqref{eq:pro4}.  
%
Then the corrected velocity $\bm U = \bm u+\bm u^*$ vanishes on the wall, at $\{x_3=0\}$. The new PDE satisfied by $\bm U$ and $P = p+p^*$ is
\begin{align}
\partial_i U_i &= \phi(\bm x)
+ \phi^*(\bm \beta \cdot \bm x) 
\label{eq:divULT}\\
-\partial_i P + \partial_m\partial_m U_i &= \psi_i(\bm x) + \psi^*_i(\bm\beta\cdot\bm x)
\label{eq:FBLT}
\end{align}
where 
\begin{align}
\phi^* &= - \phi + x_3^2 \partial_m\partial_m \phi\\
\psi_i^* &= \beta_{ij}\psi_j - 2\delta_{3i}\psi_3 + x_3^2\beta_{ij}(\partial_m\partial_m\psi_j - \partial_j\partial_m\psi_m + \partial_j\partial_m\partial_m\phi) + 2\beta_{ij}x_3(2\partial_3\psi_j - \partial_j\psi_3).
\end{align}
\end{theorem}
We discuss some consequences of the theorem before turning to its proof.  First, we note that if $\phi$ and $\bm \psi$ decay to zero in some region bounded above the wall, then $\phi^*(\bm \beta\bm x)$ and $\bm\psi^*(\bm \beta\bm x)$  vanish whenever $\bm x$ lies above the wall and the PDE satisfied by $(\bm U,P)$ is identical to that satisfied by $(\bm u,p)$.  Similarly, if $\phi$ and $\bm\psi$ do not vanish but have most of their mass above the wall, then the PDE satisfied by $(\bm U,P)$ will be modified only slightly.  

As a second consequence, we find that the Lorentz reflection procedure applied to a divergence-free $\bm u$ produces a divergence-free $\bm u^*$ (and hence $\bm U$). 
The same cannot be said for the force balance equation: if $\bm \psi=\bm 0$ we still have $\psi^* = x_3^2\beta_{ij}\partial_j\partial_m\partial_m \phi$. Therefore, the image system for the regularized point source has a nonzero force balance.  

\begin{proof}
The proof is a computation starting from the left-hand sides of \eqref{eq:divULT} and \eqref{eq:FBLT} and proceeding through unpacking of definitions to replace all instances of $\bm u$ and $p$ by $\phi$ and $\bm\psi$. Here we indicate two imprtant intermediate results. First, the biharmonic operator\footnote{The biharmonic operator appears because \eqref{eq:FBLT} contains $\nabla^2 \bm U$ and $\bm U$ depends on $\bm v$, which contains $\nabla^2\bm u$.} does not annihilate $\bm u$ but instead satisfies 
\[
\partial_k\partial_k\partial_m\partial_mu_i = \partial_m\partial_m\psi_i -\partial_i\partial_m\psi_m + \partial_i\partial_m\partial_m\phi,
\]
a result which can be recovered by examining the standard proof that $\nabla^4\bm u$ vanishes if $\phi$ and $\bm \psi$ do.  A similarly useful intermediate result is 
\[\partial_k\partial_k\partial_3u_j - \partial_k\partial_k\partial_ju_3 = \partial_3\psi_j - \partial_j\psi_3.\]
As pointed out by Kim and Karrila, the operator $(\bm u,p)\mapsto (\bm v,q)$ commutes with the operator $(\bm v,q)\mapsto (\bm u^*,p^*)$ and it is convenient to interchange these when carrying out these computations. 
We omit the rest of the argument. 
\end{proof}
We now derive the image system for the regularized point source and apply the theorem to understand exactly which PDE we have solved.  
In $\mathbb{R}^3$, the regularized point source as given by Cortez is 
\begin{align}
\begin{split}
u_i &= \frac{G'(R)X_i}{R}\\
p &= \phi(R)
\label{eq:upregfree}
\end{split}
\end{align}
where (as before) $\bm X$ denotes the vector from the source point $\bm y$ to the observation point $\bm x$ and $R = |\bm X|$. 
The function $\phi$ is a smooth function approximating the Dirac delta. More concretely, this means that $\phi = \phi_\delta(R)$ has unit mass on $\mathbb{R}^3$ and most of that mass lies within a distance of $\delta$ from the source point $\bm y$. Four possible choices of $\phi$ are listed in Table \ref{tbl:Blobs}.

 Each $\phi$ is accompanied by functions $G$ and $B$ satisfying $\nabla^2 G(R) = \phi(R)$ and $\nabla^2 B(R) = G(R)$; together $(\phi,G,B)$ is known as a \emph{blob triple}.  The flow $(\bm u,p)$ given in 
\eqref{eq:upregfree} satisfies the PDE system 
\begin{align}
\begin{split}
-\partial_i p + \partial_m\partial_m u_i = 0\\
\partial_i u_i = \phi
\end{split}
\end{align}
on all of $\mathbb{R}^3$.  To find the image system, we start with \eqref{eq:pro1}-\eqref{eq:pro2} and do some calculus to obtain
\begin{align}
\begin{split}
v_i =& 
-\beta_{ij}R^{-1}G'(R)X_j
 -2R^{-1}G'(R)x_3\delta_{i3}-2R^{-2}G''(R)X_iX_3x_3+2R^{-3}G'(R)X_iX_3x_3\\
 &+R^{-1}G'''(R)X_ix_3^{2}+2R^{-2}G''(R)X_ix_3^{2}-2R^{-3}G'(R)X_ix_3^{2}\\
q =&\;\phi(R)+2R^{-1}\phi'(R)X_3x_3+4R^{-3}G'(R)X_3^2 -4R^{-2}G''(R)X_3^2-4R^{-1}G'(R) .
\end{split}
\end{align}
We then substitute $\bm \beta\cdot\bm x$ for $\bm x$ and reflect $\bm v$ to obtain 
\begin{align}
\begin{split}
 u_i^*=&\frac{-G'(\hat{R})\hat{X_i}-2G'(\hat{R})x_3\delta_{i3}+G'''(\hat{R})\beta_{ij}\hat{X_j}x_3^2}{\hat{R}}-\frac{2G''(\hat{R})\beta_{ij}\hat{X_j}x_3y_3}{\hat{R}^2}+\frac{2G'(\hat{R})\beta_{ij}\hat{X_j}x_3y_3}{\hat{R}^3}\\
 p^*=& \phi(\hat{R})-\frac{2\phi '(\hat{R})\hat{X_3}x_3+4G'(\hat{R})}{\hat{R}}-\frac{4G''(\hat{R})\hat{X_3}^2}{\hat{R^2}}+\frac{4G'(\hat{R})\hat{X_3}^2}{\hat{R}^3}
\end{split}
\end{align}
We now pause to further simplify the expressions. One  fruitful method is to use the relations between $\phi$ and $G$ which follow from $\nabla^2G(R) = \phi(R)$, which implies the one-dimensional statement $\phi(R) = G''(R)+2G'(R)/R$.  In turn, we differentiate $R\phi = 2G'+RG''$ and rearrange to find $-2RG''+2G'  = -R^2\phi'+R^2G'''.$ This identity can be used to simplify the terms which have a factor of $\beta_{ij}\hat X_j x_3$ in $\bm u^*$:
\begin{align}
\begin{split}
u_i^* =& 
-\frac{G'(\hat{R}){X_i}}{\hat{R}}
+\beta_{ij} \hat{X}_j x_3\frac{\hat R^2 G'''\big(\hat R\big)x_3 - 2 \hat R G''\big(\hat R\big)y_3 + 2 G'\big(\hat R\big) y_3}{\hat R^3}\\
=& 
-\frac{G'(\hat{R}){X_i}}{\hat{R}}
+\beta_{ij} \hat{X}_j x_3\frac{\hat R^2 G'''\big(\hat R\big)(x_3+y_3) -  \hat R^2 \phi'\big(\hat R\big)y_3}{\hat R^3}
\end{split}
\end{align}
We proceed by canceling $\hat R^2$ and writing $-\hat X_3$ instead of $(x_3+y_3)$.  To simplify $p^*$, we again use the identity $-2RG''+2G'  = -R^2\phi'+R^2G'''$ to substitute the terms with a factor of $\hat X_3^2$ in $p^*$, and then we also note that $\hat X_3 + x_3 = -y_3$.  In the end, the formulas for the regularized point source given by Lorentz' construction are 
\begin{align}
\begin{split}
U_i =&\, \frac{G'({R}){X_i}}{{R}}
-\frac{G'(\hat{R}){X_i}}{\hat{R}}
-\beta_{ij} \hat{X}_j x_3\frac{G'''\big(\hat R\big)\hat X_3 +   \phi'\big(\hat R\big)y_3}{\hat R}\\
P =&\, \phi(R)+\phi(\hat{R})
+ 2\frac{G'''\big(\hat R\big)\hat X_3^2 - 2 G'\big(\hat R\big)
+\phi'\big(\hat R\big)\hat X_3y_3}{\hat R}
\end{split}
\end{align}
The velocity field $\bm U$ is illustrated with red arrows in the left two subplots of Figure \ref{fig:pointsource}, with source $\bm y = (0,0,1)$ and regularization length scales $\delta=\frac13,$ $\delta = \frac23$.  
The pair $(\bm U,P)$ is an exact solution of the following forced Stokes PDE system: 
\begin{align}
\begin{split}
\partial_i U_i &= \phi(R) - \phi\big(\hat R\big) + x_3^2 L\big(\hat R\big)\\
-\partial_i P + \partial_m\partial_m U_i &=
 x_3^2 \frac{\hat R^2\phi'''(\hat R)+2\hat R\phi''(\hat R) - 2\phi'(\hat R)}{{\hat R}^3}\beta_{ij}\hat X_j.
\end{split}
\end{align}
In practice, the size of this perturbation of the PDE from the free-space version depends on the decay rate of $\phi$, the regularization parameter $\delta$, and the distance $y_3$ from the source to the wall.  
\section{Novel image systems}
We used the Lorentz reflection procedure to obtain the velocities and pressures for the wall-bounded regularized Stokeslet, rotlet, stresslet, and source dipole. While in principle the rotlet, stresslet, and source dipole can be obtained from the source and the Stokeslet by differentiation with respect to $\bm y$, we found it more convenient to begin from the free-space versions using Lorentz's construction (the result is identical since differentiation with respect to $\bm y$ commutes with the transformations \eqref{eq:pro1}-\eqref{eq:pro4}). The results are listed in Table \ref{tbl:lorentz_wall_singularities} together with the results for the point source derived above. 

\begin{table}[h!]
\caption[Formulas for Green's functions in a wall-bounded domain]{
Regularized tensors of Stokes flow and the PDEs that they satisfy according to the Lorentz reflection theorem. Each velocity field vanishes on a no-slip wall at $\{x_3 = 0\}$.  Here ${\bm{X}} = \bm{x}-\bm{y}$
where $\bm{x}$ is the observation point and $\bm{y}$ is the location of the
singularity above the wall. We write $\beta_{ij} = \delta_{ij} -
2\delta_{3i}\delta_{3j}$ for the reflection operator and $\hat{X}_{i} =
\beta_{ij}x_{j} - y_{i}$. In the denominators we have written $R = |\bm{X}|$ and
$\hat{R} = | \bm{\hat{X}}|$. The function $\phi=\phi(|\bm x|)$ is a smoothed delta function, and $B$ and $G$ satisfy $\nabla^2 G(|\bm x|) = \phi(|\bm x|)$ and $\nabla^2 B(|\bm x|) = G(|\bm x|)$.  
The four scalar expressions $\lambda_i$ appearing in the Stresslet velocity and pressure are:
$\lambda_1(r) = \frac{15B'-5rG+r^2G'}{r^5}$, 
$\lambda_2(r) = \frac{-6B'+2rG-r^2G'}{r^3}$, 
$\lambda_3(r) = \frac{G'}{r} - \frac12 \phi_d$, and $\lambda_4(r) = \frac{rG''-G'}{r^3}$.  
}
\label{tbl:lorentz_wall_singularities}
{\footnotesize
\begin{tabular*}{\linewidth}{ll}
\toprule
Point source velocity 
& $ \displaystyle \Sigma_{i} =
\frac{G'({R}){X_i}}{{R}}
-\frac{G'(\hat{R}){X_i}}{\hat{R}}
-\beta_{ij} \hat{X}_j x_3\frac{G'''\big(\hat R\big)\hat X_3 +   \phi'\big(\hat R\big)y_3}{\hat R} $\\ [2mm]
Point source pressure
& $\displaystyle  \Sigma^p=
\phi(R)+\phi(\hat{R})
+ 2\frac{G'''\big(\hat R\big)\hat X_3^2 - 2 G'\big(\hat R\big)
+\phi'\big(\hat R\big)\hat X_3y_3}{\hat R}$
\\ [2mm]
Stokeslet velocity 
& $ \displaystyle S_{ij}=
\left (\frac{B'(R)}{R}-\frac{B'\big(\hat R\big)}{\hat R}-G(R)+G\big(\hat R\big)\right )\delta_{ij}-\frac{B'(R)-RB''(R)}{R^3}X_iX_j+\frac{B'\big(\hat R\big)-\hat RB''\big(\hat R\big)}{\hat R^3}(X_i\hat X_j-2x_3\beta_{ij}\hat X_3)
$
\\ [2mm]
& $ \displaystyle  {}
-\frac{B'''\big(\hat R\big)+G'\big(\hat R\big)}{\hat R}\delta_{j3}x_3\beta_{ik}\hat X_k+\frac{G'\big(\hat R\big)-\hat R\phi\big(\hat R\big)}{\hat R}\beta_{ij}x_3^2+\frac{x_3\hat R \phi \big(\hat R\big)+5\hat X_3B'''\big(\hat R\big)+3y_3G'\big(\hat R\big) }{\hat R^3}x_3\beta_{ik}\hat X_j \hat X_k
$
\\ [2mm]
Stokeslet pressure
& $\displaystyle  S^p_{j}= \frac{G'(R)X_j}{R}+\frac{2B'''\big(\hat R\big)-G'\big(\hat R\big)}{\hat R}\hat X_j+2\hat X_3 \frac{(5y_3+x_3)B'''\big(\hat R\big)-\hat Rx_3B''''\big(\hat R\big)-3y_3G'\big(\hat R\big)}{\hat R^3} \hat X_j 
$
\\ [2mm]
& $ \displaystyle {}
+\frac{4x_3B'\big(\hat R\big)-4\hat R B''\big(\hat R\big)x_3-\hat R^2(6x_3+4y_3)B'''\big(\hat R\big)}{\hat R^3}\delta_{j_3}
$
\\ [2mm]
Rotlet velocity & $\displaystyle
\mathcal{R}_{ij} = \epsilon_{mjk}\left( 
\frac{G'(R) \delta_{im}X_k }{R} - \frac{G'\big(\hat R\big) \delta_{im}\hat X_k}{\hat R}
+ x_3\hat X_k\frac{x_3\phi'\big(\hat R\big)\beta_{im}+\big(\phi'\big(\hat R\big)-G'''\big(\hat R\big)\big)\beta_{it}\hat X_t\delta_{3m}}{\hat R}+\frac{2x_3G'\big(\hat R\big) \beta_{ik}\delta_{3m}}{\hat R} \right)$
\\ [2mm]
Rotlet pressure
& $\displaystyle  \mathcal{R}^p_{j}= 0 $
\\ [2mm]
Source dipole velocity 
& $ \displaystyle D_{ij} = D_1(R)\delta_{ij}+D_2(R)X_iX_j-D_1\big(\hat R\big)\delta_{im}-D_2\big(\hat R\big)\delta_{ik}\hat X_k\hat X_m
$
\\ [2mm]
& $ \displaystyle {}
+\frac{2D_1'\big(\hat R\big)\beta_{ij}\hat X_jx_3\delta_{3m}+2D_2'\big(\hat R\big)\hat X_3 \beta_{ij} \hat X_j \hat X_m x_3}{\hat R}+2D_2\big(\hat R\big)x_3 (\hat X_m\beta_{i3}+\hat X_3\beta_{im})+D_1''\big(\hat R\big)x_3^2\beta_{im}$
\\ [2mm]
& $ \displaystyle {}
+\frac{2D_1'\big(\hat R\big)x_3^2\beta_{im}+6D_2'\big(\hat R\big)\beta_{ij}\hat X_j \hat X_mx_3^2}{\hat R}+D_2''\big(\hat R\big)x_3^2\beta_{ij}\hat X_j \hat X_m + 2D_2\big(\hat R\big)x_3^2\beta_{im}
$
\\ [2mm]
Source dipole pressure
& $\displaystyle  D^p_{j}= \frac{-\phi'(R)X_j}{R}-\frac{\phi'\big(\hat R\big)\hat X_j}{\hat{R}}-2x_3 \left(\frac{\phi'\big(\hat R\big)\hat X_3 \hat X_j}{\hat R^3}-\frac{\phi''\big(\hat R\big)\hat X_3 \hat X_j}{\hat R^2}-\frac{\phi'\big(\hat R\big)\delta_{3j}}{\hat R}\right) $
\\[2mm]
& $ \displaystyle {}
-4\left( \frac{D_1'\big(\hat R\big)\hat X_3 \delta_{3j}+D_2'\big(\hat R\big)\hat X_3 \hat X_3 \hat X_j}{\hat R}+D_2\big(\hat R\big)(\hat X_j +\hat X_3 \delta_{3j})\right)
$
\\ [2mm]
Stresslet velocity 
& $ \displaystyle S_{ijk} = 
X_i X_j X_k \lambda_1(R) 
+ \frac{\delta_{ij}X_k + \delta_{ik}X_j}{2}\lambda_2(R)
-X_i\hat X_j\hat X_k \lambda_1\big(\hat R\big) 
- 2x_3\left(y_3\beta_{ik}\hat X_j + y_3\beta_{ij}\hat X_k - x_3\delta_{jk}\beta_{im}\hat X_m\right)\lambda_1\big(\hat R\big)$
\\ [2mm]
& $ \displaystyle {}
+ x_3\beta_{im}\hat X_m\hat X_j\hat X_k\left(x_3\lambda_1''\big(\hat R\big) + (6x_3-2y_3)\frac{\lambda_1'\big(\hat R\big)}{\hat R}\right)
+\left(x_3(\delta_{3j}\beta_{ik}+\delta_{3k}\beta_{ij}) - \frac{\delta_{ij}\hat X_k + \delta_{ik}\hat X_j}{2}\right)\lambda_2\big(\hat R\big) $
\\ [2mm]
& $ \displaystyle {}
+ x_3\beta_{im}\hat X_m (\delta_{3j}\hat X_k + \delta_{3k}\hat X_j)\frac{\lambda_2'\big(\hat R\big)}{\hat R} + \frac{x_3^2}{2}(\beta_{ij}\hat X_k+\beta_{ik}\hat X_j)\left(\lambda_2''\big(\hat R\big) + 4\frac{\lambda_2'\big(\hat R\big)}{\hat R}\right)
$
\\ [2mm]Stresslet pressure
& $\displaystyle  S^p_{jk} =
\delta_{jk}\left(\lambda_3\big(\hat R\big)-2x_3\hat X_3\frac{\lambda_3'\big(\hat R\big)}{\hat R}\right)
+ \hat X_j\hat X_k \left(\lambda_4\big(\hat R\big) - 2 x_3 \hat X_3 \frac{\lambda_4'\big(\hat R\big)}{\hat R} - 4 \hat X_3^2\frac{\lambda_1'\big(\hat R\big)}{\hat R} - 4\lambda_1\big(\hat R\big)\right)$
\\ [2mm]
& $ \displaystyle {}
-2(\delta_{3j}\hat X_k + \delta_{3k}\hat X_j)\left(x_3\lambda_4\big(\hat R\big) + 2 \hat X_3 \lambda_1\big(\hat R\big) + \frac{\lambda_2'\big(\hat R\big)}{\hat R} \hat X_3\right) - 4\lambda_2\big(\hat R\big) \delta_{3j}\delta_{3k}
$
\\ [2mm]
\bottomrule
\end{tabular*}
}
\end{table}

\section{Comparison to the Ainley-Cortez expressions}
\label{sec:comp}
We undertook this work with the intention of finding a new derivation of the formulas previously given by Ainley and Cortez and their coauthors \cite{adebc08,cortez2015general}. We were surprised to find that the solutions given by Lorentz's construction are different. In this section we discuss why this does not contradict the uniqueness theory for elliptic PDEs and we give a comparison of the two versions.  

The resolution of the uniqueness question is very simple: the Lorentz and Ainley-Cortez formulas are different because they are exact solutions of forced Stokes systems with different right-hand sides. In both cases, the image systems contain regularized terms which spread up from below the wall into the fluid domain $H$, thereby perturbing either the force balance equation or both the continuity and force balance equations.  The PDE perturbations vanish in the limit of small regularization parameter $\delta \to 0$ since both systems converge to the same singular solution.  However, the perturbation and therefore the difference between the two flows become significant when $\delta$ is a significant fraction of the distance from the source point to the wall.  In Table \ref{tbl:PDEs} we list the exact PDEs for the point source and the Stokeslet in free space and in the two half-space flows. 
\begin{table}
\footnotesize
\begin{tabular}{cl}
\toprule
Flow&\quad PDEs\\
Source in $\mathbb{R}^3$&$
\begin{cases} 
-\partial_i p + \nabla^2 u_i = 0 \\
\;\quad\qquad \partial_iu_i = \phi^d(R)
 \end{cases}$\\
Source in $H$ (AC) &$
\begin{cases}
-\partial_i p + \nabla^2 u_i = 
 -2y_3\delta_{i3}{\phi^d} ''(\hat R) + 4\delta_{i3}{\hat R}
^{-1}\left(x_3\phi'(\hat R)+ y_3(\phi'(\hat R) - {\phi^d} '(\hat 
R))\right)
\\
\;\quad\qquad \partial_iu_i = 
\phi^d(R) + \phi^d\big(\hat R\big)
 \end{cases}$\\
Source in $H$ (L)&$
\begin{cases} 
-\partial_i p + \nabla^2 u_i 
\displaystyle = x_3^2 \beta_{ij}\hat X_j \frac{\hat R^2 {\phi^d}'''(\hat R) + 2{\phi^d}''(\hat R) - 2{\phi^d}(\hat R)}{\hat R^3}
\\
\;\quad\qquad \partial_iu_i 
\displaystyle = \phi^d(R) -\phi^d(\hat R) + x_3^2 {\phi^d}''(\hat R) + 2 x_3^2 \frac{{\phi^d}'(\hat R)}{\hat R}
 \end{cases}$\\
Stokeslet in $\mathbb{R}^3$&$
\begin{cases} 
-\partial_i p + \nabla^2 u_i = -f_i\,\phi(R)\\
\;\quad\qquad \partial_iu_i = 0
 \end{cases}$\\
Stokeslet in $H$ (AC) &$
\begin{cases} 
\displaystyle -\partial_i p + \nabla^2 u_i =  
-f_i\phi(R) + f_i\phi(\hat R) - 2y_3\delta_{i3}\frac{\phi'(\hat R)}{\hat R}\hat X_m\beta_{jm}f_j - y_3^2\beta_{ij}f_j \frac{\hat R{\phi^d}''(\hat R) + 2{\phi^d}'(\hat R)}{\hat R}
\\\qquad\qquad\qquad\displaystyle- 2y_3(\delta_{3i}f_j\hat X_j - f_i\hat X_3)\frac{{\phi^d}'(\hat R)-\phi'(\hat R)}{\hat R}\\
\;\quad\qquad \partial_iu_i = 0
 \end{cases}$\\
Stokeslet in $H$ (L)&$
\begin{cases} 
\displaystyle -\partial_i p + \nabla^2 u_i = -f_i\,\phi(R) - \beta_{ij}f_j\frac{\hat R\phi(\hat R) + x_3^2(\hat R\phi''(\hat R)+\phi'(\hat R)) - 4x_3\hat X_3\phi'(\hat R)}{\hat R} + 2\delta_{3i}f_3\phi(\hat R) 
\\\qquad\qquad\qquad\displaystyle+ x_3^2\beta_{ij}\hat X_j f_m \hat X_m \frac{\hat R\phi''(\hat R)-\phi'(\hat R)}{\hat R^3} - 2\beta_{ij}\hat X_j x_3 f_3 \frac{\phi'(\hat R)}{\hat R}\\
\;\quad\qquad \partial_iu_i = 0
 \end{cases}$\\
  \end{tabular}
\caption{In this table we identify the PDEs for which the regularized flows discussed here and in the work of Ainley and Cortez are exact solutions. We obtained the Lorentz PDEs from Theorem 1 and the Cortez-Varela PDEs by carefully assembling several intermediate results from that work \cite{cortez2015general}. The notation is as follows: $\bm y$ is the source point, $\bm x$ is the observation point, $\beta$ is the reflection operator, and $R = |\bm x-\bm y|$, $\hat{\bm X} = \beta\cdot\bm x-\bm y$, and  $\hat R = |\hat {\bm X}|$. 
The functions $\phi(R)$ and $\phi^d(R)$ are regularized Dirac delta functions; $\phi^d$ is known as the \emph{companion blob}. The free-space forcing function for the Stokeslet is $\bm f\phi$, and the Lorentz image system uses only $\phi$ and not a mixture of $\phi$ and $\phi^d$ as in the Ainley-Cortez-Varela construction. While the Lorentz construction for the point source could use either blob function for the forcing, we use $\phi^d$ here in order to compare to the Cortez-Varela system. } 
\label{tbl:PDEs}
\end{table}

\begin{figure}
\begin{center}
\includegraphics[width=\linewidth]{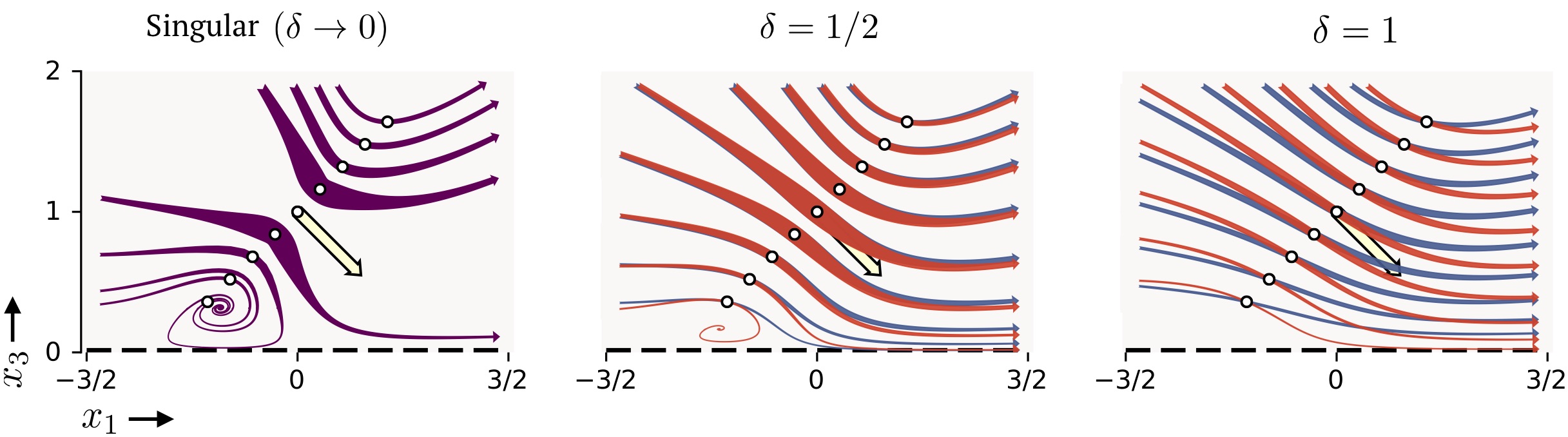}
\end{center}
\caption{Comparison of the singular wall-bounded Stokeslet and two regularized wall-bounded Stokeslets. In all three panels the source point is $\bm y=(0,0,1)$ and the direction of forcing is $\bm F=(1,0,-1)$ (pale yellow arrows).  Fluid trajectories are integrated forward and backward in time from nine seed points (white dots). All of the seed points lie on the line $x_3 = 1+x_1$ and the central one coincides with $\bm y$. Arrow widths are proportional to fluid velocity. The trajectory through $\bm y$ is omitted in the leftmost plot because the fluid velocity is infinite at that point. 
For each of $\delta=\frac12$ and $\delta=1$ we superimpose two flow fields, the Lorentz systems from this work in red and the Ainley-Cortez systems in blue. Both remain finite at $\bm y$.  The two versions mostly overlap for $\delta=1/2$ but the differences become more pronounced when $\delta=1$: the red trajectories pass closer to the wall (dashed line) than the blue trajectories. We also note the recirculation region for the Lorentz system at $\delta=1/2$, a feature which does not appear in the blue field.  The inward spiral flows in the left and center subplots do not contradict incompressibility because these are two-dimensional slices ($x_2=0$) of three-dimensional flows. 
}
\label{fig:pointforce}
\end{figure}

The two versions of the Stokeslet differ qualitatively as well as quantitatively. In Fig. \ref{fig:pointforce} we depict the singular wall-bounded Stokeslet (with $\bm f$ oriented at a $45^\circ$ angle to the wall) along with the Lorentz and Cortez flows for $\delta=1/2$ and $\delta=1$.  At $\delta=1/2$ there is only a small qualitative difference between the two flows: with the Lorentz system we see a recirculation region (similar to the singular case), while the Cortez flow has no such region.  At the higher regularization value $\delta=1$, neither flow has a recirculation. but the Lorentz flow lines approach the wall more closely.  These observations are in line with the general theme that, as the Lorentz formula avoids the use of the companion blob, its behavior resembles the singular system more than the Cortez formula using the same value of $\delta$. 

\begin{figure}
\begin{center}
\includegraphics[width=0.9\linewidth]{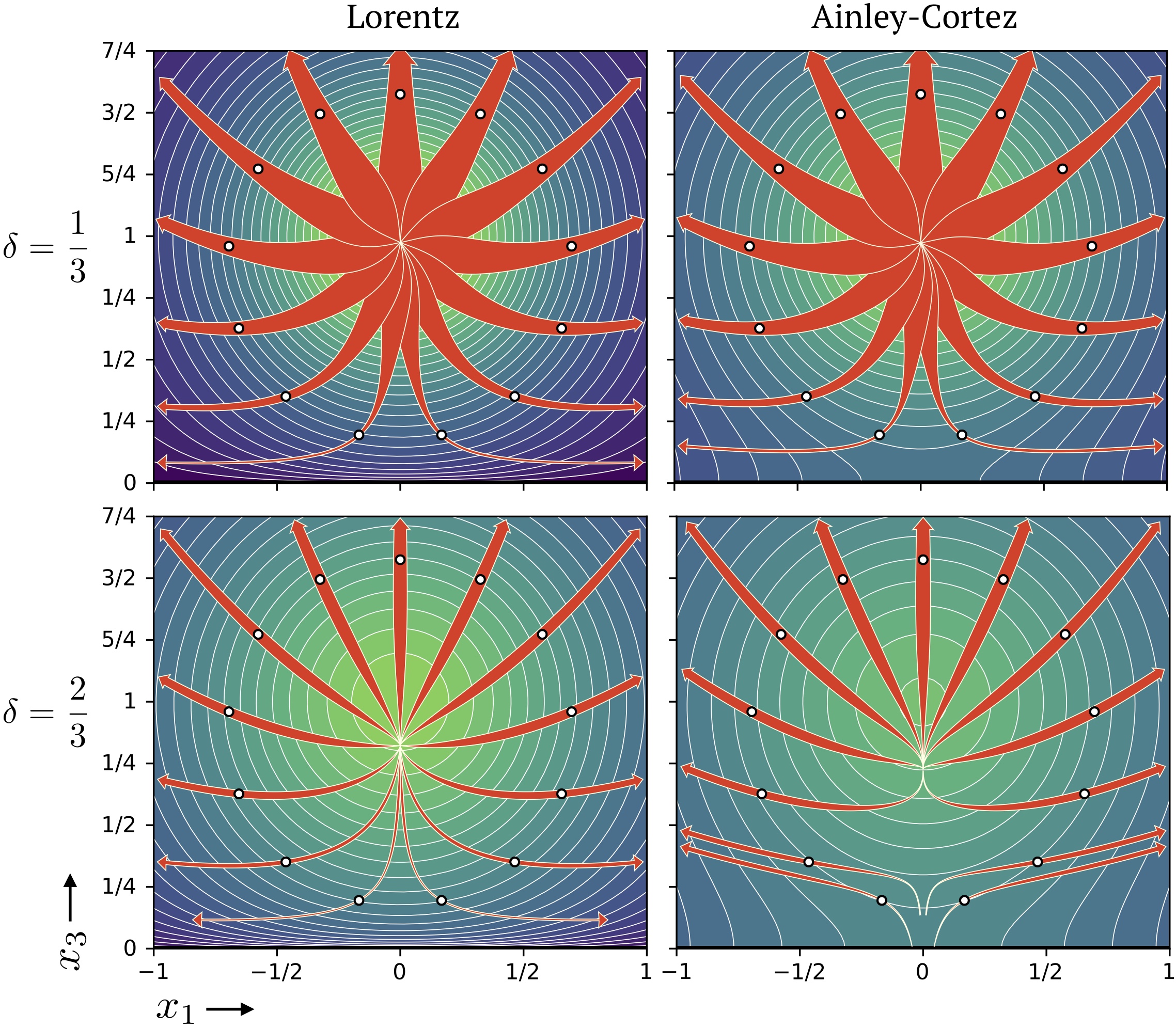}
\end{center}
\caption{Comparison of two expressions for the regularized point source near a wall: the Ainley-Cortez system and the system obtained through the Lorentz reflection theorem.  Both systems are three-dimensional flows obtained by the method of images canceling the free-space solution of the PDE $\bm \nabla\cdot\bm u = \phi^{ac}_\delta(\bm x), \-\bm\nabla p+\nabla^2\bm u = \bm 0$, where $\phi^{ac}_\delta$ is the algebraically decaying companion blob given in Table \ref{tbl:Blobs}.  We depict only the flow in the plane $\{x_2=0\}$. Both flows vanish on the wall, represented by a horizontal line at the bottom of each subplot ($x_3=0$). For small values of the regularization parameter $\delta$ (top row) the two flow fields are similar, but for larger $\delta$ (bottom row) qualitative differences emerge. The red arrows follow streamlines, with arrow thickness indicating velocity magnitude; the mapping from magnitude to arrow thickness is identical in all four subplots. The white dots indicate the seed points from which the trajectories are obtained by integration forwards and backwards in time; these lie on the circle of radius $\frac25$ and center $\left(0,\frac{7}{8}\right)$.  The contour field in the background is the divergence of velocity plotted with a logarithmic color scale.  As expected, the divergence is concentrated around $\bm y = (0,0,1)$ when $\delta$ is small, and more diffuse when $\delta$ is large. The main difference between the two systems is that the Lorentz formula has vanishing velocity as well as velocity divergence at the wall, whereas the Ainley-Cortez formula has vanishing velocity but nonzero divergence at the wall, yielding streamlines which touch the wall as $t\to-\infty$.  
}
\label{fig:pointsource}
\end{figure}

We now turn to the point source. In Fig. \ref{fig:pointsource} we depict the Lorentz and Cortez versions of the wall-bounded point source. At $\delta=1/3$ the two flow fields look very similar; at $\delta=2/3$, however, we see some of the Cortez flow lines extending backward towards the wall instead of the source point in negative time.  This is a consequence of the interesting fact that the divergence of the Cortez velocity field does not vanish on the wall.  

To compare the two flows more quantitatively, we define errors in both systems using volume integrals. We focus on the point source and the Stokeslet because the other singularities are all constructed from derivatives of these two.  First, let $\bm u$ and $p$ be the velocity and pressure for the wall-bounded regularized point source so that $\bm \nabla\cdot\bm u(\bm x) \approx \phi^d(|\bm x-\bm y|)$ and $-\bm\nabla p + \nabla^2\bm u \approx 0$.  For both the Ainley-Cortez and the Lorentz systems we define the errors
\begin{align}
E_1 &= \int_H \left|\bm\nabla\cdot\bm u - \phi^d\right|\,dV_{\bm x} + \int_H \left|-\bm\nabla p + \nabla^2 \bm u\right|\,dV_{\bm x}\label{eq:E1}\\
\displaystyle E_2 &= \left|1-\int_H \left|\bm\nabla\cdot\bm u\right|\,dV_{\bm x}\right|\label{eq:E2}.
\end{align}
The first error $E_1$ measures the PDE perturbation induced by the forcing inherent in the image systems. The second error $E_2$ measures the deviation from unity of the integral of the velocity divergence over the fluid domain.  We note that it is impossible for both $E_1$ and $E_2$ to vanish if $\phi^d(|\bm x-\bm y|)$ has any support below the wall: if $E_1=0$ then the velocity divergence matches $\bm \phi^d$ perfectly, but this implies $E_2>0$.  Therefore, some error is inevitable for any method based on radially symmetric, non compactly supported regularizations of the delta function.  Turning to the regularized wall-bounded Stokeslet, we suppose now that $\bm \nabla\cdot\bm u\approx 0$ and $-\bm\nabla p + \nabla^2\bm u \approx -\bm f \phi$.  In this case, we define the errors  
\begin{align}
E_3 &= \frac{1}{\left|\bm f\right|}\left(\int_H \left|-\bm\nabla p + \nabla^2 \bm u + \bm f \phi(\bm x-\bm y)\right|\,dV + \int_H\left|\bm\nabla\cdot\bm u\right|\,dV\right) \label{eq:E3}\\
E_4 &= \frac{1}{\left|\bm f\right|}\left| \bm f + \int_H \left(-\bm\nabla p + \nabla^2 \bm u\right)\,dV\right|
\label{eq:E4}
\end{align}
In fact, both the Ainley-Cortez and Lorentz versions of the Stokeslet are exactly divergence-free, so the second integral in \eqref{eq:E3} could be omitted.  
As with the point source, the first error integral measures the PDE perturbation induced by the forcing of the image systems while the second error integral in some sense measures the deviation from unity of the mass of a regularized delta function.  Both $E_3$ and $E_4$ are normalized by the length of the force vector $\bm f$.  

\begin{figure}
\[\includegraphics[width=0.8\linewidth]{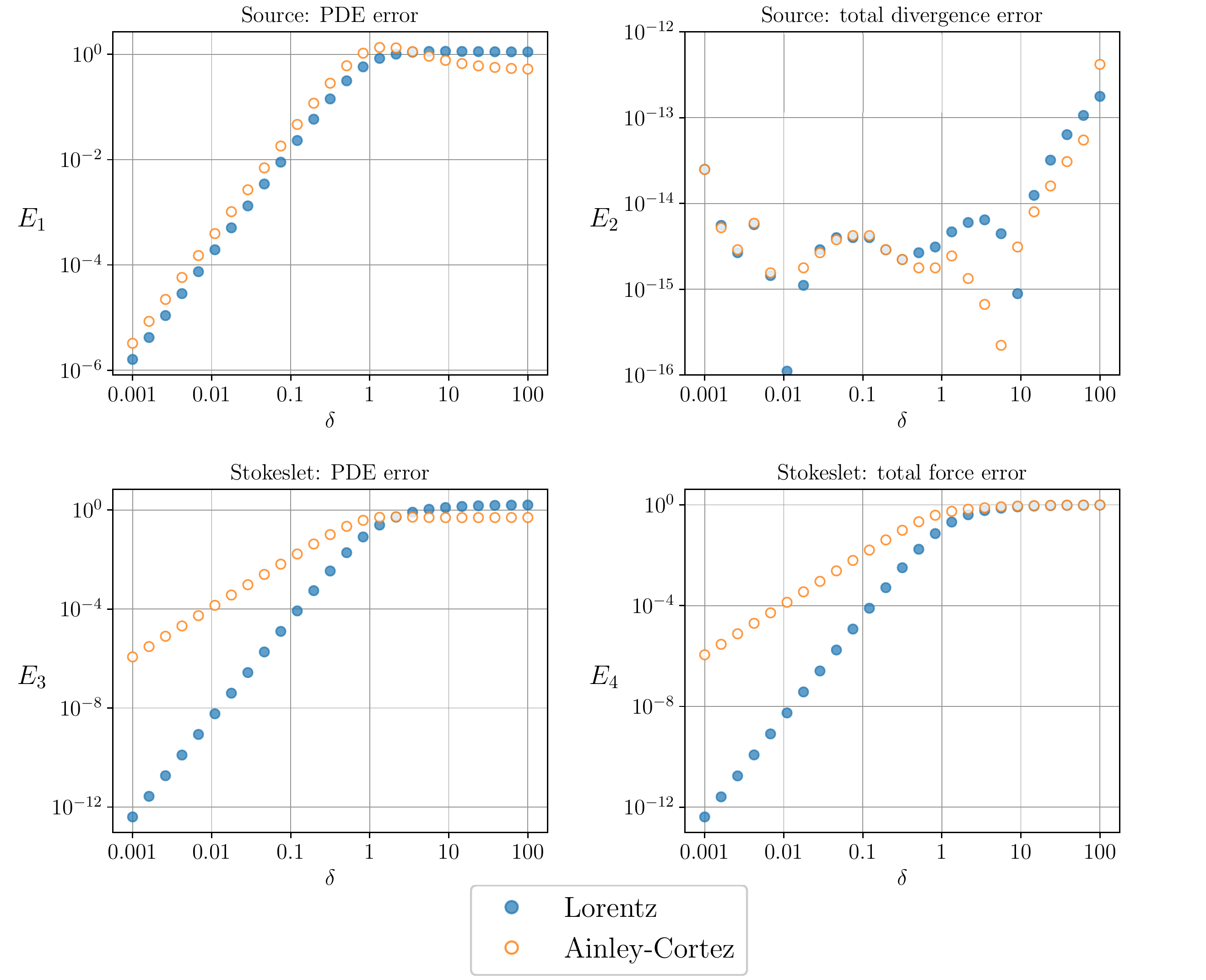}\]
\caption{Some errors are inevitable in canceling systems built from radially symmetric and non-compactly supported blob functions. In all cases, we place a regularized free-space source or Stokeslet above the wall at $\bm y=(0,0,1)$ and cancel the flow along the wall at $\{x_3=0\}$ using either a Lorentz or a Cortez image system at $-\bm y$. We plot the four errors defined in equations \eqref{eq:E1}-\eqref{eq:E4} against the regularization parameter $\delta$.  When $\delta$ is large, the forcing terms in the image system are able to spread out into the fluid domain $H = \{\bm x:x_3>0\}$; in contrast, at small values of $\delta$ the forcing in the image systems is well localized behind the wall, leading to smaller errors. 
The point source PDE error $E_1$ is similar for the Lorentz and Ainley-Cortez image systems, although the Lorentz systems are better by a factor of about 2 for small $\delta$.  
For the Stokeslet errors $E_3$ and $E_4$, we see different asymptotic error rates: proportional to $\delta^4$ for the Lorentz systems and $\delta^2$ for the Ainley-Cortez systems.  This reflects the fact that the Ainley-Cortez image system for the Stokeslet is built in part from a more slowly decaying companion blob, whereas the Lorentz version is not. On the other hand, the Cortez systems have slightly smaller errors with $\delta\gg y_3$. The error $E_2$ is analytically zero with the Cortez system, so the top right panel demonstrates the accuracy of our numerical integration procedure (it also suggests that $E_2$ may vanish for the Lorentz system as well, although we do not prove this). 
These computations suggest that the Lorentz version of the wall-bounded Stokeslet may be preferable to the Ainley-Cortez version, although both versions should be used with caution when the regularization parameter $\delta$ is on the order of the distance from the source point to the wall. }
\label{fig:E1234}
\end{figure}

We carried out numerical integration to evaluate the four error functions $E_1$-$E_4$ at values of the regularization parameter ranging from $\delta = 10^{-3}$ to $\delta = 3$.  
To evaluate these integrals over $H$ we used iterated Gauss integration in spherical coordinates centered at $\bm y =(0,0,1)$:
\begin{equation}
\int_H  \psi(\bm x)\,dV = \int_0^{\pi} \int_0^{U(\phi)} \int_0^{2\pi} \psi(r,\phi,\theta) r^2\sin\phi\,d\theta\,dr\,d\phi
\end{equation}
The upper limit of the radial coordinate is $U(\phi)=\infty$ for $\phi<\pi/2$ and $U(\phi)=\sec(\pi-\phi)$ for $\phi>\pi/2$.  Infinite integration intervals in $r$ are transformed to finite intervals in $u = (1+r)^{-1}$.  
The results are given in Fig. \ref{fig:E1234} (using the algebraically decaying blobs $\phi^a$ and $\phi^{ad}$).  We find that the point source errors $E_1$ and $E_2$ are similar between the Ainley-Cortez systems \cite{cortez2015general} and the Lorentz systems derived here; for small $\delta$, the Lorentz error $E_1$ is an improvement by a factor of 2. For the Stokeslet system, we see a more dramatic improvement, with $E_3$ and $E_4$ proportional to $\delta^4$ for the Lorentz systems and proportional to $\delta^2$ for the Ainley-Cortez systems. 


In order to evaluate the errors $E_3$ and $E_4$, we required an expression for the pressure of the Cortez regularized half-space Stokeslet. After some study of the paper that gave a formula for the velocity \cite{cortez2015general}, we wrote down the following expression:
\begin{equation}
p(\bm x) = X_iF_i\frac{G'(R)}{R}
-\beta_{ij}\hat X_j F_i\frac{G'(\hat R)}{\hat R}
+ 2f_3y_3\frac{G'(\hat R)}{\hat R}
- 2 y_3 \hat X_iF_i(x_3+y_3)\frac{\hat R G''(\hat R)-G'(\hat R)}{\hat R^3} 
+ y_3^2 \hat X_iF_i\frac{{\phi^d}'(\hat R)}{\hat R}
.\end{equation}
Here $\bm x$, $\bm y$, $\bm X$, $\hat{\bm X}$ and so on are as defined above (not in the notation of \cite{cortez2015general}).

\section{Conclusion}
We have presented a novel set of image systems which cancel the regularized Green's functions of Stokes flow on a plane wall. These new image systems are the result of applying Lorentz's reflection theorem to the free-space regularized systems given by Cortez, Fauci, and Medovikov \cite{Fauci2005}, but they differ from the image systems derived in previous works \cite{adebc08,cortez2015general}. The numerical integration of four types of errors suggests that the Lorentz versions will be advantageous in much, but not all of the parameter space defined by the source height $y_3$ and regularization parameter $\delta$. In practice, we have found that the sensitivity of the method of regularized Stokeslets to the choice $\delta$ is a more important numerical consideration than choice between the Lorentz and Cortez versions of the image systems. We plan to address these numerical issues in a future work. 

\bibliographystyle{plainnat}
\bibliography{MyLibrary}

\begin{thebibliography}{6}
\providecommand{\natexlab}[1]{#1}
\providecommand{\url}[1]{\texttt{#1}}
\expandafter\ifx\csname urlstyle\endcsname\relax
  \providecommand{\doi}[1]{doi: #1}\else
  \providecommand{\doi}{doi: \begingroup \urlstyle{rm}\Url}\fi

\bibitem[Ainley et~al.(2008)Ainley, Durkin, Embid, Boindala, and
  Cortez]{adebc08}
J.~Ainley, S.~Durkin, R.~Embid, P.~Boindala, and R.~Cortez.
\newblock The method of images for regularized {S}tokeslets.
\newblock \emph{J. Comput. Phys.}, 227:\penalty0 4600--4616, 2008.

\bibitem[Cortez et~al.(2005)Cortez, Fauci, and Medovikov]{Fauci2005}
R.~Cortez, L.~Fauci, and A.~Medovikov.
\newblock {The method of regularized Stokeslets in three dimensions: Analysis,
  validation, and application to helical swimming}.
\newblock \emph{Phys. Fluids}, 17:\penalty0 031504, 2005.

\bibitem[Cortez and Varela(2015)]{cortez2015general}
Ricardo Cortez and Douglas Varela.
\newblock A general system of images for regularized stokeslets and other
  elements near a plane wall.
\newblock \emph{Journal of Computational Physics}, 285:\penalty0 41--54, 2015.

\bibitem[Kim and Karrila(1991)]{kk91}
S.~Kim and S.~J. Karrila.
\newblock \emph{Microhydrodynamics: {P}rinciples and {S}elected
  {A}pplications}.
\newblock Butterworth-Heinemann, Boston, 1991.

\bibitem[Kuiken(1996)]{kuiken1996ha}
H.~K. Kuiken.
\newblock {HA Lorentz: Sketches of his work on slow viscous flow and some other
  areas in fluid mechanics and the background against which it arose}.
\newblock \emph{J. Engr. Math.}, 30:\penalty0 1--18, 1996.

\bibitem[Lorentz(1896)]{lorentz1896}
H.A. Lorentz.
\newblock Eene algemeene stelling omtrent de beweging eener vloeistof met
  wrijving en eenige daaruit afgeleide gevolgen.
\newblock \emph{Zittingsverslag van de Koninklijke Akademie van Wetenschappen
  te Amsterdam}, \penalty0 (5):\penalty0 168--175, 1896.

\end{thebibliography}

\end{document}